\documentclass{article}

\usepackage[utf8]{inputenc}
\usepackage{authblk}
\usepackage{tabularx}
\usepackage{graphicx}
\usepackage{booktabs}
\usepackage{hyperref}
\usepackage{natbib}
	\setcitestyle{authoryear,round,aysep={}}
	
\providecommand{\keywords}[1]{\textbf{\textit{Keywords: }} #1}

\begin{document}
\title{VIDA: A simulation model of domestic VIolence in times of social DistAncing}

\author[1,2]{Lígia Mori Madeira}
\author[2,3]{Bernardo Alves Furtado}
\author[1,2]{Alan Rafael Dill}

\affil[1]{Department of Political Science at Federal University of Rio Grande do Sul (UFRGS)}
\affil[2]{National Council for Technological Development (CNPq)}
\affil[3]{Institute for Applied Economic Research (Ipea)}

\maketitle   

\begin{abstract}
Violence against women occurs predominantly in the family and domestic context. The COVID-19 pandemic led Brazil to recommend and, at times, impose social distancing, with the partial closure of economic activities, schools, and restrictions on events and public services. Preliminary evidence shows that intense coexistence increases domestic violence, while social distancing measures may have prevented access to public services and networks, information, and help. We propose an agent-based model (ABM), called VIDA, to illustrate and examine multi-causal factors that influence events that generate violence. A central part of the model is the multi-causal stress indicator, created as a probability trigger of domestic violence occurring within the family environment. Two experimental design tests were performed: (a) absence or presence of the deterrence system of domestic violence against women and (b) measures to increase social distancing. VIDA presents comparative results for metropolitan regions and neighbourhoods considered in the experiments. Results suggest that social distancing measures, particularly those encouraging staying at home, may have increased domestic violence against women by about 10\%. VIDA suggests further that more populated areas have comparatively fewer cases per hundred thousand women than less populous capitals or rural areas of urban concentrations. This paper contributes to the literature by formalising, to the best of our knowledge, the first model of domestic violence through agent-based modelling, using empirical detailed socioeconomic, demographic, educational, gender, and race data at the intraurban level (census sectors). 
\end{abstract}

\keywords{Domestic Violence, Violence against Women, Agent-based Models (ABMs), Pandemics, Simulation, Metropolitan Regions}

\section{Introduction}

Domestic violence is a global public health and human rights violation issue \citep{opas_folha_2017}. It is estimated that around 30\% of women in the world suffer, or have suffered violence, usually committed by intimate male partners. Many cases result in homicides, with almost 40\% of female murders committed by their partners. UN Women estimates that in countries like France, Cyprus, Singapore, and Argentina, social distancing has increased cases of domestic violence by 25\% to 30\% \citep{un_women_headquarters_covid-19_2020}. A report by WHO and PAHO also denounces that domestic violence against women tends to increase in times of emergency, including epidemics, victimising more vulnerable female groups such as the elderly, women with disabilities, refugees, and those living in conflict-affected areas \citep{who_covid-19_2020}.

Under normal circumstances, this panorama reveals the seriousness with which the problem must be addressed. During the pandemic with the recommendations for isolation and social distancing, authorities, media, and researchers have pointed to an increase in domestic violence rates in many countries. This is also the case in Brazil, if not worse \citep{warken_assassinatos_2020}. Data released by the National Ombudsman for Human Rights, office linked to the Ministry of Women, Family, and Human Rights, indicate that calls to the hotline grew 14\%, with an increase of 37.6\% in April compared to the same month in 2019 \citep{antunes_violencia_2020}.

Social distancing alters the internal dynamics of domestic violence against women by intensifying the factors at the root of this violence, such as gender inequality, the patriarchal system, the macho culture, and misogyny. Also, domestic violence may increase in the context of the pandemic due to the consequent economic impact, the overload of reproductive labour on women, stress and other emotional effects, alcohol or drug abuse, and the reduction of helplines and services \citep{alencar_politicas_2020}.

This study seeks to illustrate situations of domestic violence against women before and after the COVID-19 pandemic using agent-based simulation modelling, reproducing the main findings in the literature regarding the causes of domestic violence. The model proposed (called VIDA) contributes to understanding the mechanisms behind the hypothesis that staying at home as a social distancing measure affects the magnitude of numbers related to domestic violence. Our experimental design introduces deterrence and social distancing measures in the model, to compare with the baseline model that reflected the context before the pandemic. 

The simulation demonstrated that the difference in the effects of deterrence in each cohort due to the availability of the criminal justice system and informal control networks (community) was the mechanism guiding the explanation for the increase in cases of domestic violence against women while following social distancing measures. Specifically, the study estimated metrics for the economic decline and for the behaviour of staying at home as a social distancing measure that allows comparing Brazilian metropolitan regions. Therefore, the simulation could grasp regional differences while contributing to understanding the main factors around social distancing and economic decline, as well as the existence and features of deterrence measures.

Section two provides an overview of the literature on the causes of domestic violence committed by intimate partners. Section three describes the methodology and presents the model VIDA. We explain the model’s idea, assumptions, and implementation based on the construction of the indicator of stress and the characterisation of the artificial population. Also, we discuss the model’s validation, calibration, and tests of policies that illustrate domestic violence, before and during the application of social distancing measures. Section four shows the general results and analysis, based on weighted areas, for each of the Brazilian metropolitan regions. Section five closes this article, presenting our final considerations.

\section{Domestic violence and femicides: a brief literature review}

The literature review on partner abuse perpetration \citep{rothman_chapter_2018} lists several causes and determinants of this type of violence. The discussion starts from the definition of the very concept of dating abuse, now understood as physical, sexual, verbal, and emotional abuse directed at a partner (present or ex), occurring in relationships of people of the opposite or same sex \citep{center_for_disease_control_preventing_2019}, not distinguishing between the type of abuse, its severity, the consequences of these acts, the frequency with which they occur, or differentiating the gender of perpetrator and victim. This concept does not include violence characterised as being chronic, severe, resulting in physical or sexual injury, caused by a person in a position of advantage over the victim. These types generally characterise the domestic violence against women and are taken by social movements as a core element in their definition, given the characteristic of power and control, involving what the literature calls patriarchal terrorism, of the abuser towards the victim.

Based on a sociological perspective, Johnson (\citeyear{johnson_patriarchal_1995}) demonstrates two different ways to approach the issue of domestic violence. One is the family violence perspective, in which the object refers to family conflicts due to their frequency, the role of stress, and the adherence to normative standards that would accept some form of violence in a family context. The other is the feminist perspective, which restricts the focus to specific factors of the perpetration of violence against women by intimate partners, prevailing the historical, cultural, and social patterns of patriarchy. Other concepts observed in the literature are intimate terrorism, a form of violence aimed at controlling the relationship; and situational violence between partners, which does not result from patterns of abuse and control but escalates to violence in the emergence of specific situations of conflict \citep{johnson_differential_2005}.

According to Stark (\citeyear{stark_looking_2012}), the conceptualisation and, consequently, how to address domestic violence, are divided into two models, the “violent incident model” and the “coercive control model”. The first defines violence between partners as a criminal offence since the late 1990s. It considers measures such as restricting access to victims, children, and firearms, and is based on shelter programs, legal assistance, and other support for victims. In the violent incident model, incarceration is the preferred response for offenders, and the level of danger observed is the main factor determining how the police respond to domestic abuse. Abusers who continue their actions – with sufficient time between each event – are considered, by analogy to criminals in general, as repeat offenders. Despite the contribution to reducing severe forms of violence, scholars question whether incarceration and judicial protection can have a long-term impact.

The coercive control model is defended as a more comprehensive approach. It considers that most female victims of violence are subjected to a pattern of dominance that includes techniques of isolation, degradation, exploitation, control, and physical violence. It is a pattern of psychological and emotional abuse that the literature calls patriarchal or intimate terrorism \citep{johnson_typology_2008}, or coercive control \citep{stark_coercive_2019,stark_coercive_2020}. Stark shares that “this gap between what the law defines as the crime of domestic violence and the actual tactics abusers use to subjugate their partners severely limits the efficacy of even the most dedicated and well-trained police” (\citeyear[p. 201]{stark_looking_2012}). For advocates of this model, it is necessary to change from a pattern based on specific incident response to a proactive response that redefines partner abuse as ongoing conduct. It is crucial to apply appropriate sanctions to stop the course of such conduct and the risk of escalation. Studies have advanced in understanding the risk to children living in environments of domestic violence \citep{katz_beyond_2016}. Despite criticism, scholars argue that police intervention in events of domestic violence has prevented thousands of injuries and deaths, in addition to changing the normative pattern regarding violence against women \citep{stark_looking_2012}.

The concern with causal explanations for violence between intimate partners began in the last century, involving different mechanisms, ranging from psychoanalytic theories, explanations of intergenerational transmission, and social learning, to theories of situational background. Bell and Naugle (\citeyear{bell_intimate_2008}) review the main theories of intimate partner violence (IPV) with variables of interest and theoretical limitations. The main explanations encompass (a) the idea that frustration in one area of life can lead to aggression – for example, low autonomy in the workplace could lead to abusive compensation at home –, (b) the assumption that childhood and the environment where a person was raised play a role in predicting future abusive partners, and (c) the theories mixing explanations of distal background, such as childhood abuse, exposure to interparental aggression, and situations that trigger violence (alcohol consumption, jealousy, and other conflicts with the partner). The authors propose a framework that, based on different theories, offers a contextual analysis of IPV including distal and proximal aspects, proposing that the perpetration of abuse can be explained by antecedent factors, motivators, unwritten rules, and legal and social consequences of violent behaviour \citep{bell_intimate_2008}. According to the authors, the advantage of this integrating theoretical framework is the flexibility in presenting common points between apparently contradictory findings.

Sociocultural theories range from economic explanations to approaches of structural inequality (present in explanations of feminist theories). Models based on inequality consider that domestic abuse reflects the power structure in society, taking into account income, unemployment, culture, race, and gender. Feminist theories share the understanding that an unequal society influences the emergence of incidents of violence. They emphasise, however, that male privilege overcomes any other social inequality in a society with a patriarchal structure. Feminist theories originally proposed that the primary reasons for domestic violence were related to a reinforcement of gender standards, but later adjustments suggested the need to accommodate inter-gender relations. Because of the prevalence of feminists in the political coalitions that formulated programs against this type of violence, these initiatives tend to assume that gender norms and traditional roles lead to perpetration and victimisation.

Criminology theories have taken advantage of recent developments in the study of urban crime and criminal “hot spots” and have relied on the routine activity theory of criminal events and on the explanations related to target or victim suitability to understand domestic and intimate violence \citep{mannon_domestic_1997}. These approaches consider crime as predatory, involving a motivated offender, an available target, and the absence of capable guardians. Family isolation also contributes to increasing victimisation, making the home one of the most dangerous places for women and children.

Finally, socio-ecological models organise multiple determinants at different levels (individual, family, peer, community, institutional, and societal). Violent behaviour cannot be attributed to a single factor but to a wide range of possible grouped factors, including biological and psychological risk factors, childhood experiences, economic status, substance use, social characteristics (friends, for instance), aspects of the intimate relationship with the partner, workplace culture, and level of involvement with education, public policies, disadvantaged neighbourhood, violence in the community, access to weapons, cultural norms related to aggression, the woman’s social status, intersectional nature, and multiple forms of oppression \citep{rothman_chapter_2018}.

More recent studies have distinguished the lethal risk of IPV from domestic and family violence (DFV). The prediction of each of these types of violence has different numbers and causes. For Ferguson and McLachlan (\citeyear{quteprints199780}), many tools to access risk are designed to predict violence and its escalation, including re-victimisation and repeated offences, not necessarily to predict homicides. While IPV is consistently identified as one of the high-risk factors for femicides, the risk factors may not be the same, for example, for the use and abuse of drugs and alcohol. The prevalence of drugs and alcohol abuse is much higher in cases of physical violence than in the prediction of femicides \citep{quteprints199780}. On the other hand, having a gun increases the risk of femicide by 1000\%, but such a risk would not be confirmed in cases of domestic violence. Also, the characteristics of the relationship, the presence of children, marital status, and the duration of the relationship do not appear to be significant risk factors for femicides, although they may contribute to the risk of intimate violence.

\section{Methodology}

This section explains the concept of agent-based models (ABMs) and describes the proposed model, VIDA. After this introductory part, subsection 3.1 will address the model’s intuition, followed by subsection 3.2 that describes the generation of the indicator of stress. Subsection 3.3 explains the experimental design and the proposed changes in parameters related to measures of social distancing and deterrence of domestic violence. Finally, subsection 3.4 presents the model’s sensitivity analysis and validation.

Agent-based models (ABMs) are built by describing agents (individual and active subjects) and the interrelationship between them, as well as the environment where these interactions occur \citep{epstein_growing_1996}. ABMs are artificial simulations made in a computational environment. They mimic the core mechanisms of a phenomenon under examination \citep{abdou_designing_2012}. Therefore, ABMs are simulacra, artificial, in silico, reduced to core aspects of the phenomenon, and their purpose is well-defined. 

Epstein and Axtell (\citeyear{epstein_growing_1996}) synthesise ABMs as the analysis of the functions of transformation of agents and environment, as a result of the present interaction between the agents and the environment. These transformation algorithms, the sequence, and how interactions occur, are formally described and can be verified and reproduced through the computational code provided, which is considered a best practice \citep{grimm_odd_2010,grimm_odd_2020}. ABMs are subject to verification and scrutiny of the background theory, and the processes and empirical data adopted.

Understanding criminal processes is an area of criminology in which ABMs can be useful and contribute to the construction of public policies and the action of public managers. Among the aspects of criminology, opportunity theories investigate the motivations of aggressors to understand the environmental contexts in which crimes occur. When developing models to test these opportunities, it is necessary to predict the dynamic interactions of individuals involved in each criminal event, their interactions with other agents and the environment.

A criminal system is guided by a wide range of interrelated factors. These factors include and are not limited to the individual perception of the offender, the configuration and knowledge of the physical environment, the convenience and attractiveness of the target or victim, the cognitive representation of the environment, and other factors related to the community.

The criticisms and importance of ABMs result from the difficulty of criminology theories hitherto addressing the importance of individual incidents located in specific time and space. The theories have always been concerned with general and aggregate standards, making it difficult to draw conclusions regarding the behaviour of victims and offenders that could affect crime rates and occurrence. It is necessary to examine the individual actors who play essential roles in criminal events in order to understand the patterns and characteristics of crimes better.

\subsection{The Intuition of the Model VIDA}

In normal circumstances, the reality of families encompasses individual members participating in professional or social activities, most of the time outside the home. During a pandemic where society is upstanding social distancing measures, family reality changes, and individuals are at home most of the time, even when carrying out work activities. Environmental and socioeconomic factors continue to contribute to this experience, intensifying social and pathological factors such as the use and abuse of drugs and alcohol.

Regarding the issue of domestic violence against women, it is noted that the deterrent mechanisms are compromised in a context of social distancing by at least two aspects. First, it is hard or even impossible to be distant from the abuser. Second, there are fewer opportunities to denounce the abusive situation to the community, as well as less access to the police and reduced response by the justice system. 

The purpose of the model VIDA is to illustrate causal elements intrinsic to the domestic and family environment that favours the likelihood of violence against women. It contributes to the increase in the theory’s explanatory capacity and offers quantitative elements to support the validity of the causal elements embedded in the model. When the general behaviour of conditions for violence is calibrated according to the figures available for Brazil, it is possible to construct indicators of the magnitude of adverse effects resulting from a specific contextual situation. In this study, VIDA explicitly tested the increase in families’ permanence at home due to social distancing measures taken to prevent the spread of the virus Sars-CoV-2 in the context of the COVID-19 pandemic and the restriction to deterrence systems available to women victims of domestic abuse.

The model is based on the following family situation: male abuser and female victim, with or without children. The characteristics of men, women, and the family environment are derived as samples of the population as observed in the census of 2010 conducted by the Brazilian Institute of Geography and Statistics (IBGE), for Brazilian metropolitan regions. Based on the family nucleus, we build a risk indicator (stress) that accumulates additively \citep{berge_unique_2014} theoretical hypotheses that contribute as a trigger to domestic violence. This risk indicator serves as a probabilistic factor that leads to violence.

\subsection{Construction of the indicator of stress from familial and model context}

The indicator of stress is the central part of the model and aims at systematically weighing in the findings of the literature. The indicator is created successively adding: (a) exogenous characteristics of each family member (model’s agents), (b) endogenous agents' variables that accumulate throughout the simulation, and (c) variables chosen by the modeller. 

Aspects related to income, gender, years of education, age, race, and the typical size of families are exogenously included. The variables are collected from census sectors and weighted statistical areas. Endogenous variables result from other variables or the interaction promoted by the model’s algorithm. The modeller may change some variables to assess the model’s response. These variables include the proportion of those staying at home (before and during the pandemic), the relevance of gender to the indicator of stress, level of employment, the possession or not of weapons, and presence of substance use disorder.

The starting point to establish the indicator of stress is gender, considering that being male leads to a higher indicator (see table \ref{tab:1}). After that, we add individual, household and per capita income influence, specifying that lower income leads to comparatively higher indicators. Household income works as a proxy for neighbourhood quality and income per capita as a proxy for domestic (dis)comfort. 

The literature review in section two further suggests four hypotheses related to the composition of the indicator of stress. The variable years of schooling is added to the indicator based on the hypothesis that fewer than six years of study leads to a higher indicator of stress of 60\%, and the greater the number of years of schooling, the lower the indicator. Another hypothesis is that being older than 18 and younger than 29 years old is a characteristic relevant to the indicator of stress. The third hypothesis concerns the victim’s race. In this case, black agents receive a proportional increase of 30\% in the indicator of stress. Finally, the fourth hypothesis leads to an increase in the indicator of stress for women who are employed.

Some factors undergo multiplicative proportionality of the value to be added, according to their relevance (see table \ref{tab:1}). Those considered highly relevant are multiplied by ten, and medium relevant, by five. For years of schooling (which ranges from 0 to 17) and history of domestic abuse (number of events of violence), values are divided by a constant (10). These values were chosen to maintain a sensible balance among factors influencing the indicator of stress and were validated via the sensitivity analysis performed.

Access to firearms and substance use disorder are included as aggravating factors. Firearms are considered twice as highly relevant. Access to firearms, once confirmed, is deterministic (always added to the indicator). Substance use disorder, in turn, is implemented as a random additional factor because it may or may not occur.

As such, the indicator of stress is the result of the addition and multiplication of a number of composition individual and familial census characteristics that follow existing literature reasoning and degree of relevance. The composite measure is then used probabilistically as a violence trigger that accumulates endogenously. This tentative construction captures the subjective findings of the literature in a systemic way and enables our implementation of a deterrence system mechanism.  

\begin{table}[!t]
\begin{tabular}{lllll}
\toprule
Indicator of stress       & Values added & Observations & Source     & Relevance\\
\midrule
Gender                    & Male 0.8, female 0.2    & Initial value & Modeller & \\
Income                    & 1 -- income           &        & IBGE 2010 & High  \\
Household income          & -- household income    &      & IBGE 2010 & Medium \\
Income per capita         & 1 -- income pc &       & IBGE 2010 & Medium\\
Yrs. of schooling    & 1 -- (schooling / 10)     & + 60\% if \textless 6 & IBGE 2010 & High\\
Age                       & 1, If \textgreater 18 age \textless 29  &  & IBGE 2010 & High\\
Spouse’s race      & 1, if black female         &        + 30\%           & IBGE 2010 &\\
Employment & 1, if employed         & Def.: 80\% pop.     & Modeller   & Medium       \\
Staying at home           & 0.67, if no work; 0.34    & & Endogenous & Medium       \\
Firearms        & 1, if access; 0 & Def.: 10\% pop.       & Modeller   & High * High  \\
Substance use     & 1, if addicted; 0 & Def.: 10\% pop.   & Modeller   & High, random \\
Hist. domestic abuse & Events violence / 10   & & Endogenous & High        \\
\bottomrule
\end{tabular}
\caption{The table lists the components of the indicator of stress, how they affect it, the source of the data and the relevance of the multiplicative proportionality.}
\label{tab:1}
\end{table}

VIDA is a model with simple dynamics. In each simulation, VIDA starts by generating the artificial population sample according to the choices made by the modeller. The model then runs for a brief period of 10 steps. During this period, there is some stochastic neutral volatility associated with employment and income. Moreover, there is enough time for an endogenous history of violence to build up. After 200 steps, the violence levels have stabilized and the model is interrupted and output is generated. The model also includes a fixed model scale that divides the indicator by a 1,000 so to calibrate the cases of attacks to the available empirical data. The execution of each step of the model consists of:

\begin{enumerate}
    \item Update of the indicator of stress for all agents.
    \item Verification of the violence trigger.
    \item Verification of the need to seek help from public policies against violence, the deterrence process.
\end{enumerate}

The full model is available at the \href{https://www.comses.net/codebases/ade80a69-82fa-4258-88ae-d17fb7094e23/releases/1.0.2/}{COmSES platform} and \href{https://github.com/BAFurtado/VIDA_domestic_violence}{GitHub}. It is built upon the Mesa framework \citep{masad_mesa:_2015}. The reader interested in using VIDA should focus on the 'violence' folder. Basically, the construction of the stress indicator is included as an agent step at \texttt{agents.py} and the creation of the sampled families from data is made within \texttt{model.py}, using data and processes from the folder 'input'. \texttt{Generalization.py} iterates the model 200 times and tests the experimental design. Outputs are also included in the repository.

\subsection{Experimental design: social distancing and deterrence of domestic violence}

When the model appropriately represents the phenomena researched, it is possible to ask what-if questions. Therefore, it is possible to test the likely effects of interventions that have not yet taken place. Our experimental design includes two consequences of the pandemic: (a) the imposition of social distancing measure of staying at home, and (b) the access (or lack thereof) of the victim to protective measures in case of domestic violence (deterrence system). 

Victims searching for help when suffering from domestic violence typically have three similarly-sized behaviour patters, according to data compiled by the Brazilian Senate. The first group is of victims that never denounce. Victims from the second group in contrast denounce immediately after the first case of violence. Victims from the third group denounce after the third event of violence. Following this segmentation, the victims in the model VIDA are distributed equally among these three groups and act accordingly. 

The model then considers that women who denounce have a 50\% chance of obtaining protection. Among those who denounce and obtain protection, 50\% have the chance to see their abuser convicted. The victim’s situation (denounce, protection, or abuser conviction) is included in the model as an element that reduces the abuser’s indicator of stress (see table \ref{tab:4}. Therefore, it works to reduce the chances of further violence cases, and each request for support in the deterrence system reduces the likelihood of future violent events.

\begin{table}[!t]
\centering
\begin{tabular}{llll}
\toprule
Indicator of stress       & Values added           & Source     & Relevance    \\
\midrule
Denounce                   & -- 1       & Endogenous   &     Medium \\
Protective measure        & -- 1         & Endogenous & High         \\
Conviction      & -- 1          & Endogenous & High       \\
\bottomrule
\end{tabular}
\caption{Influence of deterrence system on aggressor's indicator of stress.}
\label{tab:4}
\end{table}

\subsection{Comments on calibration, validation and sensitivity analysis}

VIDA is simulated with different sampled data. The results presented refer to the median of results obtained after the 200-fold simulation. The model was calibrated in order to approximate the number of notifications of domestic violence against women made by health agencies in 2011, using data from the 2010 census. In 2011, data from the Senate indicated 73.7 notifications per 100,000 women.

The literature offers different concepts of model validation \citep{galan_errors_2009,guerini_method_2017,moss_alternative_2008,ngo_calibration_2012,wilensky_introduction_2015}. However, there is some consensus around the idea that models are validated according to their purpose, context and ontology \citep{edmonds_simulating_2017}. A model proposed for predicting is expected to offer accurate and confirmed results that were not part of the model beforehand. A model designed to help to understand a given mechanism does not have to prove a predicted empirical data. In any case, a model must be anchored in its purpose and reflect the available literature. 

Accordingly, VIDA's purpose is to illustrate causal elements that are subjectively present in the literature, thus helping understand the phenomena and enable the implementation of what-if analysis. As a first tentative to systematically represent a multi-causal intricate theoretical mechanism -- that of domestic violence trigger -- and the empirical foundation of the input of the model, we believe VIDA results are sufficiently validated to enable comparisons among metropolitan regions and insights into intraurban differences. 

The sensitivity analysis results presented in the next section contributes to demonstrate the robustness of the model as it assesses how the variation of parameters and processes interfere in the results. Moreover, the analysis of parameter changes contributes to comprehending the effects these changes generate.

The modeller running VIDA can control the variation of the parameters that are not endogenous or derived directly from the IBGE population sampling. Thus, the modeller may alter data on: relevance of gender; employment percentage (and consequently the population staying at home and whether receiving salaries); access to firearms, and substance use disorder. It is possible to test volatility of employment and income value, increasing the chances of moving from employed to unemployed and vice versa, as well small increases or decreases in income.

\section{Results}

First, we present the results of the sensitivity analysis, where the parameters chosen by the modeller are submitted to variation. Second, we present the baseline results and those for the experimental design. Then, we show a comparison of results across 46 metropolitan regions in Brazil. Finally, we detail intraurban results for Brasília and Porto Alegre. Once the model was calibrated with nationwide data from 2011, showing 73.7 reports of violence per 100,000 women, it was possible to observe the differences resulting from the implicit variation of race, gender, age, family size, presence of weapons, and economic capacity of families in different metropolitan regions and intrametropolitan neighbourhoods. 

\subsection{Sensitivity analysis}

The results presented adopt the standard parameters of the model for the metropolitan region of Brasília (see table \ref{tab:2}), with the presence of deterrence and absence of social distancing measures, that is, the situation before the pandemic. The model runs 200 simulations for each tested parameter. 

The indicator of stress according to gender, fixed at 0.2 for women and varying for men between 0.1 and 0.9, has little influence on the results related to increasing cases and denounces. The percentage increase between the lowest and highest value was 2.24\% for cases of violence and 2.48\% for denounces.

The percentage of employment between 0.1 and 0.9, led to an increase of 4.85\% in domestic violence against women, and an increase of 5.36\% in the number of denounces. This slight increase is in accordance with the suggestion of the literature that working women are more subject to domestic violence. 

As expected, the introduction of a high * high relevance multiplicative proportionality for access to firearms into the model resulted in an increase in cases and denouncing. Varying from 0.1 to 0.9, access to firearms increased violence cases significantly by 72.43\% and denouncing by 66.52\%.

The occurrence of cases of domestic violence related to substance use disorder included an additional random factor, considering that such cases may or may not occur. Thus, the increase (from 0.01 to 0.5) in the chances of being addicted resulted in an increase in violence cases of 7.44\% and the number of denounces of 7.86\%.

\begin{table}[!t]
\centering
\begin{tabular}{lll}
\toprule
Indicator component level                   & Cases  & Denounces \\
\midrule
Gender stress                &        &           \\
0.1                          & 169.57 & 60.05     \\
0.44                         & 172.28 & 61.72     \\
0.9                          & 173.45 & 61.58     \\
\midrule
Percentage of employed  &        &           \\
0.1                          & 165.66 & 58.65     \\
0.44                         & 169.14 & 60.81     \\
0.9                          & 174.11 & 61.98     \\
\midrule
Firearms                     &        &           \\
0.1                          & 173.33 & 62.4      \\
0.44                         & 368.05 & 115.29    \\
0.9                          & 628.79 & 186.41    \\
\midrule
Substance use disorder       &        &           \\
0.01                         & 170.99 & 60.94     \\
0.22                         & 178.09 & 63.68     \\
0.5                          & 184.73 & 66.14    \\
\bottomrule			
\end{tabular}
\caption{Sensitivity analysis: results of cases and denounces per 100,000 women when varying modeller choices of indicator os stress components.}
\label{tab:2}
\end{table}

\subsection{Simulation results}

The model VIDA with typical parameters (see table \ref{tab:1}) was simulated 200 times to report the average numbers obtained. The average number of denounces was 61.73, in an environment with 174.49 domestic violence cases per 100,000 women.

Table \ref{tab:3} shows the results for the experimental design tests. Social distancing measures, which forces the agents to stay at home, regardless of whether they are working, leads to an increase in violence cases of 8.86\%. At the same time, denounces reduced by almost a third (29.25\%), compared with the baseline simulated model with the default parameters.

Additionally, the inclusion of the deterrence system in the model demonstrated a reduction in domestic violence cases by only 3.53\%, when compared with its absence (178.89 versus 172.79 cases per 100,000 women). These findings corroborate the literature \citep{campbell_risk_2003}, which states that judicial protection and incarceration have a greater long-term impact and work to reduce severe forms of violence, especially femicides. According to the literature, other aspects contribute to reducing the risk of severe forms of domestic violence, such as the expansion of employment, preventing substance use disorder, and restricting access to weapons.


\begin{table}[!t]
\centering
\begin{tabular}{llll}
\toprule
Deterrence & Social distancing & Cases  & Denounces \\
\midrule
FALSE      & FALSE             & 180.38 & 0         \\
FALSE      & TRUE              & 197.35 & 0         \\
\midrule
TRUE       & FALSE             & 173.62 & 61.97     \\
TRUE       & TRUE              & 190.74 & 47.38    \\
\bottomrule
\end{tabular}
\caption{Results per 100,000 women for different alternatives of social distancing and deterrence system.}
\label{tab:3}
\end{table}

\begin{figure}[!t]
\centering
\includegraphics[width=\textwidth]{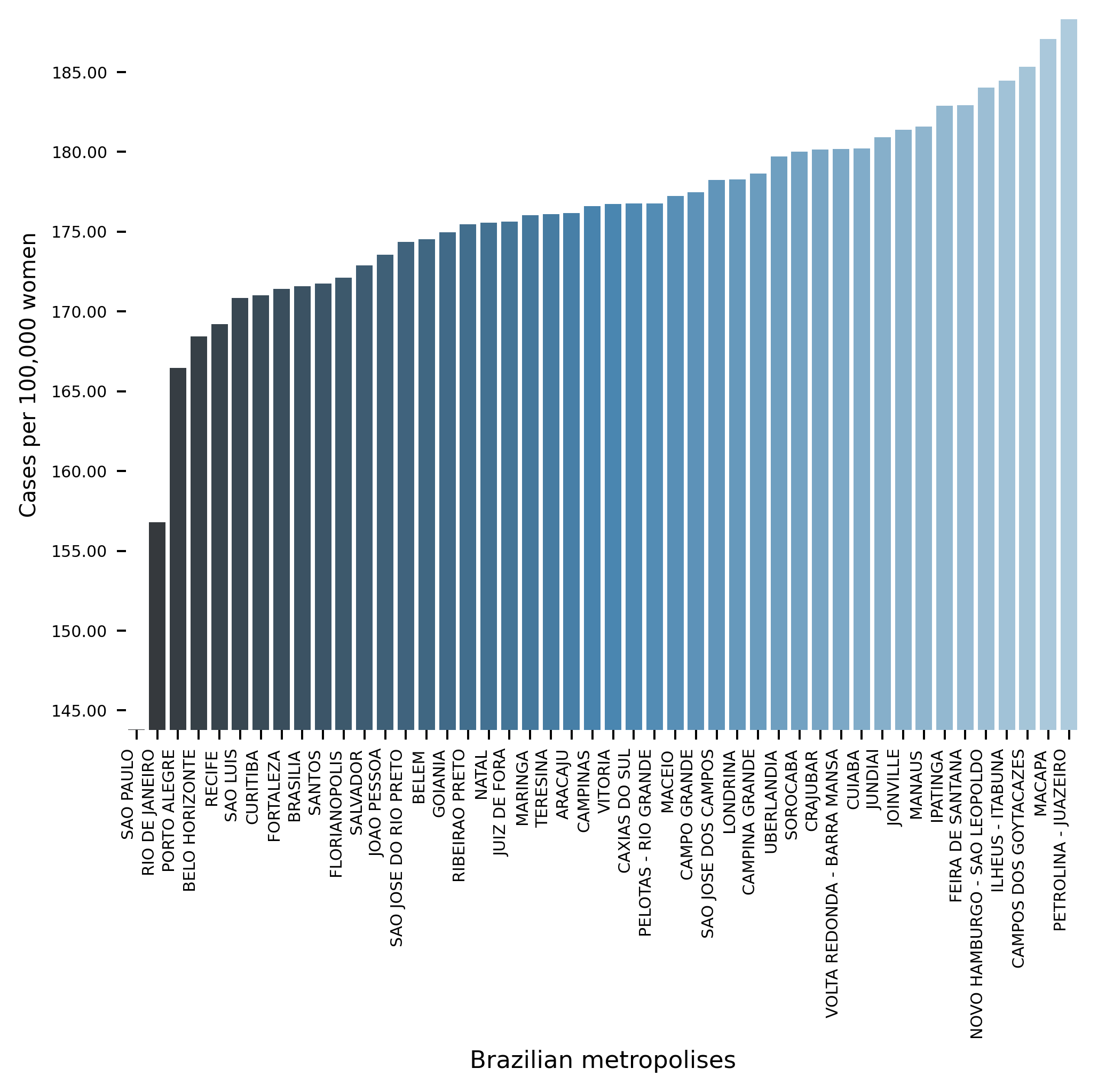}
\caption{Number of domestic violence cases per 100,000 women in Brazilian high-populated areas metropolitan regions as classified by the IBGE.}
\label{fig:1}
\end{figure}

A relevant contribution of the simulation is the ability to compare the results across different metropolitan regions with varying composing attributes, such as age and gender, family size, average wages, years of schooling, and race distribution, according to the 2010 demographic census sample data. Figure \ref{fig:1} shows the number of cases of domestic violence per 100,000 women in the high-populated areas in Brazilian metropolitan regions. The data simulated in the model suggest that larger and wealthier metropolitan regions have proportionately fewer cases of domestic violence against women than regions with smaller and poorer populations.

Empirical analysis for Brazil have shown increasing rates in the regions North and Northeast, and decreasing rates in the South and Southeast of Brazil in the past decade. When shifting the focus of analysis to the size of the municipalities, there is a general decrease in homicides in large cities (with more than 500,000 inhabitants) and an increase in medium-sized (between 100,000 and 500,000 inhabitants) and small municipalities (with less than 100,000 inhabitants) \citep{cerqueira_atlas_2019}. Despite their sizes, another important explanatory element that may affect the results is the lack of infrastructure and deterrence systems in regions with a larger number of domestic violence cases \citep{madeira_capacidade_2018}. Our simulation results reveal patterns that are similar to the empirical data, showing notably fewer cases in metropolitan regions of the South and Southeast. When cases are more frequent in these regions, they are identified in metropolitan regions that encompass smaller cities, located inland and with less access to deterrence apparatus.

\begin{figure}[!t]
\centering
\includegraphics[width=\textwidth]{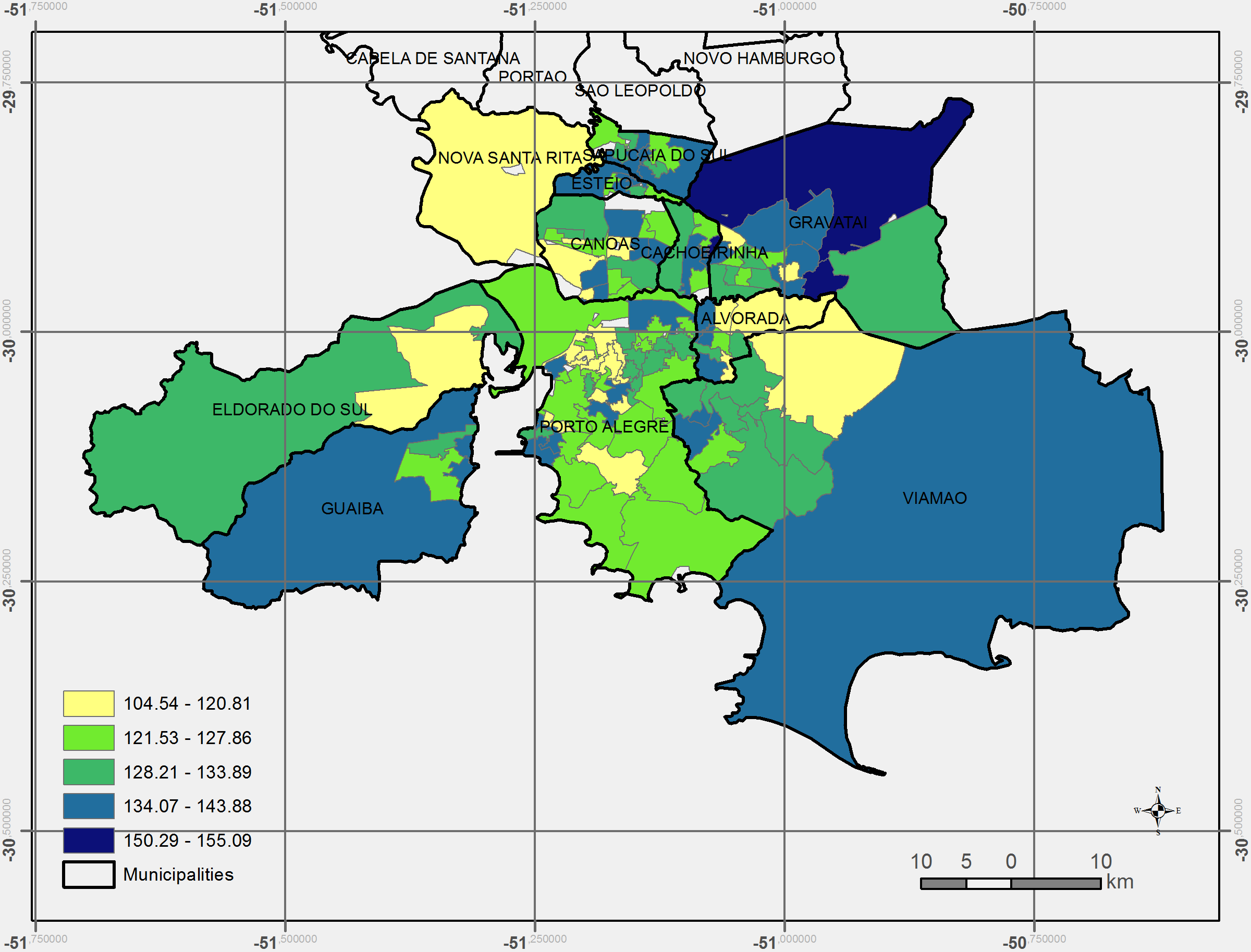}
\caption{Cases per 100,000 women. Mean obtained from 500 simulations for domestic violence cases per 100,000 women in weighted areas (IBGE), for the metropolitan region of Porto Alegre}
\label{fig:2}
\end{figure}

Further, we compared intraurban simulation results with the current reality (see figure \ref{fig:2}). The comparison data was retrieved from the database of Indicators of Violence Against Women — Maria da Penha Law, for the year 2019. \footnote{Retrieved from https://www.ssp.rs.gov.br/indicadores-da-violencia-contra-a-mulher} The analysis from the municipalities located in the metropolitan region of Porto Alegre \footnote{Retrieved from https://atlassocioeconomico.rs.gov.br/regiao-metropolitana-de-porto-alegre-rmpa} suggest that results show similar patterns of domestic violence when comparing the simulation and reality. Neighbourhoods corresponding to municipalities with high rates of violence, such as Viamão, Taquara, Cachoeirinha and Guaíba figure in the simulation among the groups with the highest incidence. Alvorada, with a proven record of violence, appears in the simulation with an western portion that is in the second tier of highest violence, along with an eastern, more sparsely inhabited area that is less violent. The small municipality of Nova Santa Rita which is comprised by a single neighbourhood (statistically weighted area) shows a dispersed average and stands out from a simulated low number of attacks, as compared to intermediate levels in the obtained data.


\begin{figure}[!t]
\centering
\includegraphics[width=\textwidth]{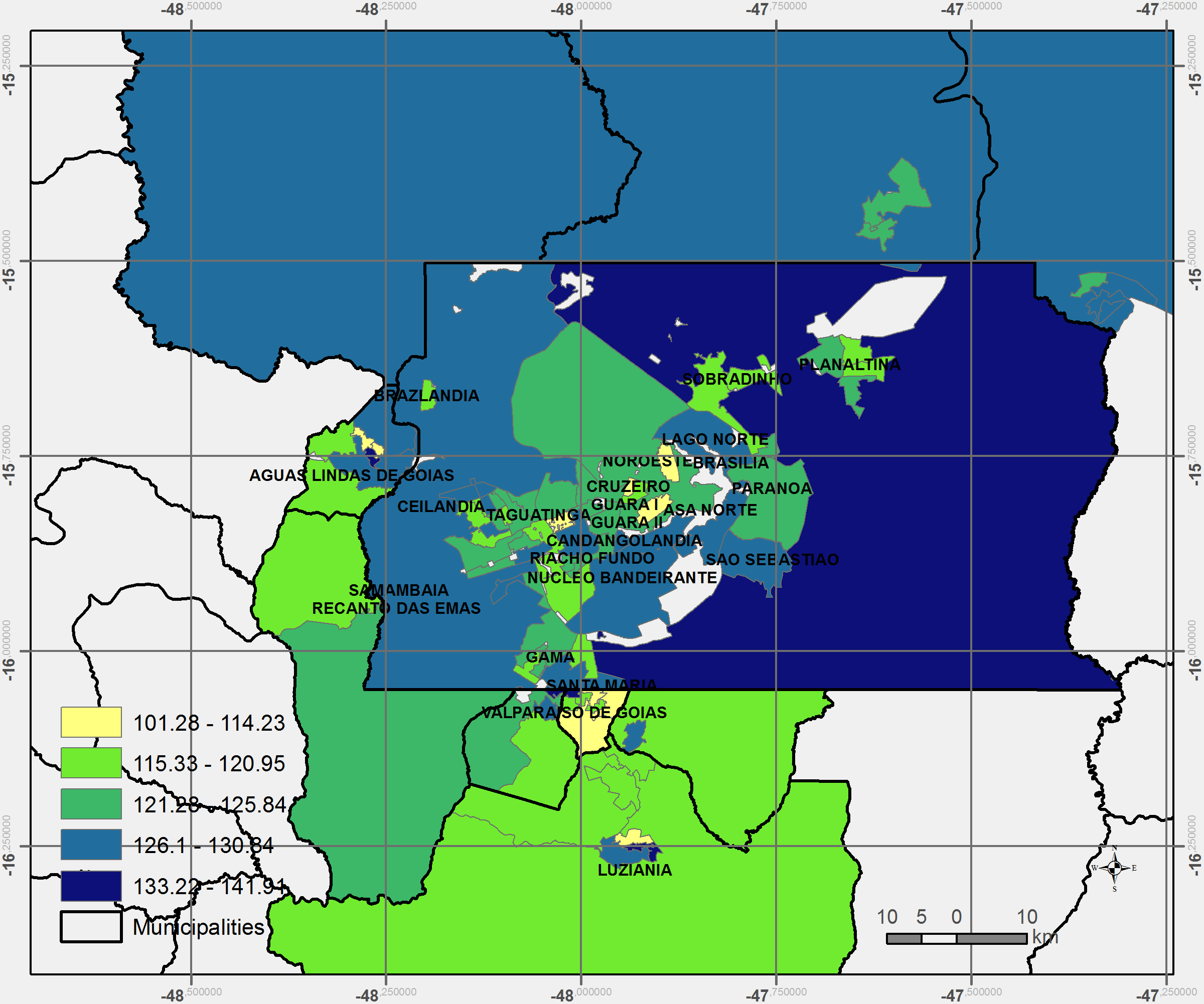}
\caption{Cases per 100,000 women. Mean obtained from 500 simulations for domestic violence cases per 100,000 women in weighted areas (IBGE), for the metropolitan region of Brasilia.}
\label{fig:3}
\end{figure}

Figure \ref{fig:3} shows the violence cases in the metropolitan region of Brasília (Federal District). In this case, it appears that the eastern, very large and sparsely populated rural portion of the region and some peripheral urban centres are more susceptible to violence against women than in the most populated areas. Our results show neighboring areas of high violence, side by side with less violent areas out of the Federal District bounds, in peripheral areas of the south, namely Valparaíso de Goías and Luziânia, and Águas Lindas de Goiás to the west. Those results suggest the possibility of high intraurban distinctions of violence that are not captured by municipalities averages, but that may be corroborated by police hot-spot analysis. 

This exercise can be replicated for all metropolitan regions. The adoption of real data allows studying specific weighted areas and simulating the geospatial distribution of domestic violence cases, which contributes to the formulation of policies to address this issue.

\section{Final considerations}

This article presents a first attempt to formalise a multi-causal model of domestic violence, based on a series of family characteristics and comparable artificial data, generated from intra-urban (large neighbourhoods) census data. The model illustrates theoretical mechanisms and quantifies explanatory aspects of violence against women. Additionally, it incorporates aspects that emerged with the COVID-19 pandemic – forced coexistence in the home environment due to social distancing measures – and elements of the deterrence system to address domestic abuse. As such, the paper contributes to the understanding of the complex mechanisms encompassing violence against women in Brazil.

The results suggest an increase in violence against women of about 10\% with the implementation of social distancing measures emphasising staying at home, which is confirmed by recent data that point to a 30\% decrease in denounces in several countries and 25\% in some Brazilian states \citep{bueno_violencia_2020}. More densely populated metropolitan regions appear to have fewer violence cases per 100,000 women compared to smaller metropolitan regions. The same comparative pattern appears in the intra-metropolitan analysis between high-populated areas and rural or peripheral areas.

In addition, the model, its codification, and the process of generating artificial population samples based on census data are publicly available. This practice aims to facilitate the dissemination of the model and its use by the scientific community, analysts, and public policy managers.

Future work involves detailing empirical data in the model, specifically identifying possession of firearms per household, and feedback effects within the neighbourhood. Data on the access of firearms can greatly contribute to the reliability of the proposed spatial analysis because the presence of firearms seems to be the strongest among the causes of femicides.

The elaboration of public policies depends on the right diagnoses capable of supporting appropriate interventions. The area of violence and crime routinely deals with the problem of lack of data, even more so in cases of high underreporting such as domestic violence. The ethical limitations of producing experiments in the area can also be overcome by using agent-based simulating models, which make it possible to replicate reality and illustrate different effects in an auspicious way. However, researchers who have been trying to understand the risks related to domestic abuse, especially in the context of the pandemic, have suggested the need to use data parsimoniously since it takes time to understand the changes in the dynamics of collecting and recording information. The model demonstrated to be useful to simulate and anticipate data, allowing to illustrate this type of violence and its georeferenced disparities. This tool proves to be of great importance in this particular moment where all kinds of inequalities stand out in the country.










\bibliographystyle{apalike}
\bibliography{vida} 


\end{document}